\documentclass[11pt,twoside]{article}

\usepackage{asp2004}
\usepackage{epsf}
\usepackage{psfig}
\usepackage{lscape}
\usepackage{graphicx}

\newcommand{\msun}{\ensuremath{\mathit{M}_{\odot}}}                  % solar masses: msun
\newcommand{\lsun}{\ensuremath{\mathit{L}_{\odot}}}                  % solar luminosity
\newcommand{\zsun}{\ensuremath{\mathit{Z}_{\odot}}}                  % solar metal content

\newcommand{\teff}{\ensuremath{\mathit{T}_{\rm eff}}}                % effectieve temperatuur
\begin{document}
\title{Spectral Synthesis of Massive Stars in Clusters}    %%% Fill in title
\author{Claus Leitherer}   %%% Fill in author names
\affil{Space Telescope Science Institute, 3700 San Martin Drive, Baltimore,
    MD 21218~$^1$}
\altaffiltext{1}{Operated by AURA, Inc., for NASA under contract NAS5-26555}

\begin{abstract} %%% Abstract to run on from here.
Stellar clusters are thought to be the simplest stellar systems and the closest observational counterparts to theoretical models for single stellar populations. Progress in our understanding of the atmospheres and evolution of massive stars has led to generally reliable synthesis models. The future release of new evolution models with rotation, however, will require non-trivial updates to previously published synthesis models, in particular for all Wolf-Rayet and red supergiant related quantities. Cluster synthesis work is currently progressing from a purely stellar approach to a more comprehensive stellar+cluster perspective. The photometric evolution of stars and the dynamical evolution of clusters are delicately interwoven. Recent work attempts to combine these seemingly related fields.
\end{abstract}

%%% MAIN BODY OF TEXT GOES HERE. CONSULT "INSTRUCTIONS FOR AUTHORS USING
%%% LATEX2E MARKUP", SECTIONS 2.3-2.6 FOR HELP WITH EQUATIONS, FIGURES,
%%% AND TABLES.

\section{Introduction}

Population synthesis aims to reproduce integrated properties of arbitrary complex stellar systems by numerically  superposing the properties of individual stars. Such properties can be spectral energy distributions, spectral line profiles, or mass and energy return. A variety of tools and methods exist to tackle this problem (e.g., Fioc \& Rocca-Volmerange 1997, Leitherer et al. 1999, Cervi\~no, Mas-Hesse, \& Kunth 2002, Schulz et al. 2002, Bruzual \& Charlot 2003, Robert et al. 2003, V\'azquez \& Leitherer 2005). Massive hot stars are of particular interest, as their intrinsic luminosity makes them a dominant contributor to the total light output of a population. In addition, populations of
massive stars at the largest explored redshifts are now accessible to direct observation (Steidel et al. 2003), providing a further incentive for synthesis modeling.

Star clusters are thought to be the simplest stellar population systems, as opposed to, e.g., entire galaxies with their complex star formation histories. Naturally, clusters have become the preferred training sets for testing synthesis models. An example is in Fig.~1, where the recent synthesis models of Gonz\'alez Delgado et al. (2005) are compared to three LMC clusters of young, intermediate, and old ages. The theoretical and observed spectra are in excellent agreement. Age-sensitive tracers such as the Ca H+K/H$\epsilon$ features are precise chronometers. Nevertheless, when viewed at high spatial resolution, almost any bona fide simple star cluster turns out to be a more complex aggregate of different stellar generations. 30~Doradus, the archetypal starburst cluster, hosts a previous stellar generation of age $>$~10~Myr, the current 2~--~3~Myr old starburst, and a new, dust obscured stellar generation found in the infrared (Walborn \& Barb\'a 1999, Grebel \& Chu 2000).

In this review I will first give a brief overview of the state of the art in synthesis modeling, with an emphasis on hot, massive stars. I will point out recent improvements in modeling that are afforded by progress in stellar atmospheres. At the same time, significant shortcomings still exist, some of which may be resolved once the new generation of evolution models with rotation becomes available. The final part of this review attempts to address the combined effects of stellar and cluster evolution. This subject has become a recent favorite of Henny Lamers and is blossoming both in terms of theoretical breakthroughs and observational data of unprecedented quality.

\begin{figure}[h]
\begin{center}
\includegraphics[scale=.50,angle=-90]{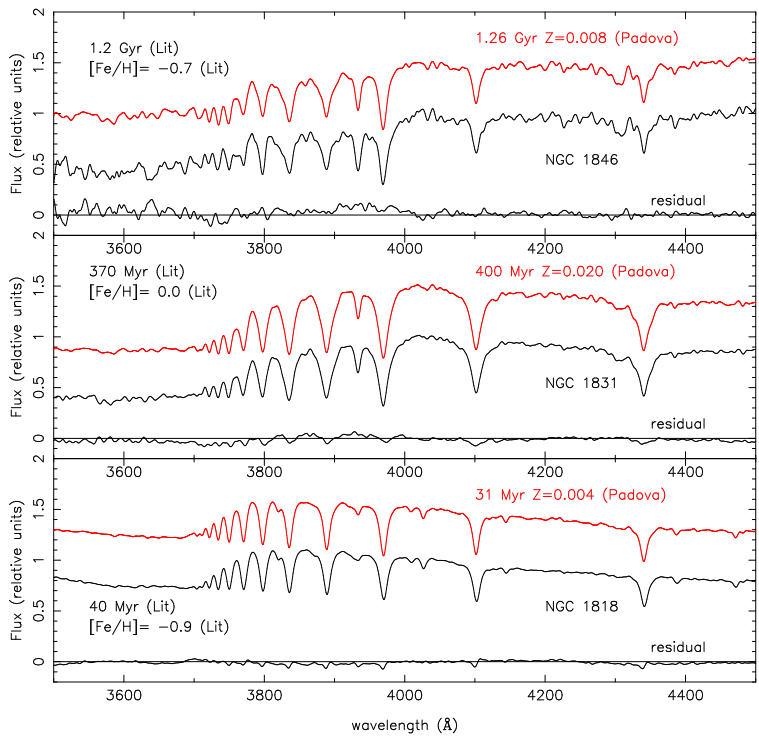}
\end{center}
\caption{Intermediate resolution spectra of three LMC clusters, NGC 1818 (bottom), NGC 1831 (center) and NGC 1846 (top), are compared with synthetic spectra. The metallicity and age reported in the literature and in the models are indicated. The spectra are flux normalized at 4010~\AA. From Gonz\'alez Delgado et al. (2005).}
\end{figure}

\section{Parameter Space}

I will focus on extragalactic star clusters whose sizes are small enough to permit integrated light studies. One of HST's major contributions to observational astronomy was the discovery of young star clusters in galaxies with current star formation (Holtzman et al. 1992). A large fraction (whose precise value is under debate [Lada \& Lada 2003]) of the newly formed stars are located in compact star clusters. These clusters are found to have radii of a few pc, ages of a few to hundreds of Myr, and masses between a few hundred to more than $\sim$10$^8$~\msun\ (Whitmore 2003). Typically, clusters are $\sim$10~Myr old and have a mass of $\sim$10$^5$~\msun. These objects are the targets of most spectral synthesis studies.

\begin{figure}[h]
\begin{center}
\includegraphics[scale=.60]{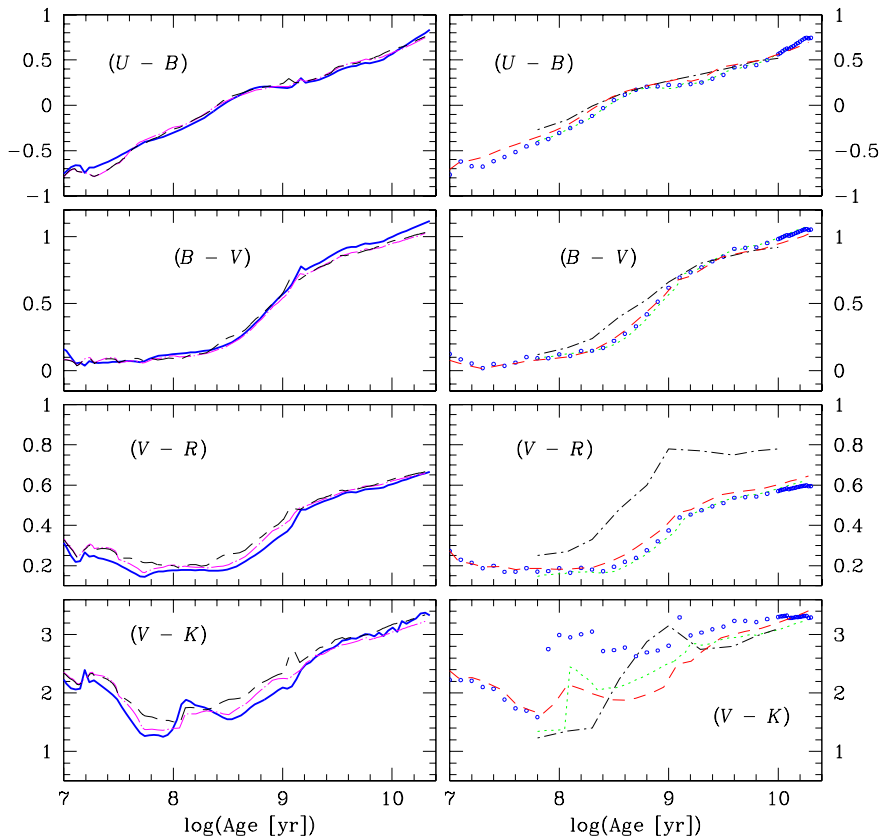}
\end{center}
\caption{Predicted color evolution for different models with solar composition. Thick solid line: V\'azquez \& Leitherer (2005); long-dashed+short-dashed line: Bruzual \& Charlot (2003) with Padova 1994 tracks; dotted+long-dashed line: Bruzual \& Charlot with Padova 2000 tracks; open small circles: V\'azquez, Carigi, \& Gonz\'alez (2003) and Padova 1994 isochrones; dotted line: V\'azquez, Carigi, \& Gonz\'alez and Padova 2000 isochrones; dot+short-dashed line: Maraston (1998); short-dashed line: Fioc \& Rocca-Volmerange (1997). The curves were distributed over the left and right panels to improve clarity. From V\'azquez \& Leitherer (2005).}
\end{figure}

The large observed age and mass spread results from a combination of selection effects and physical processes. Older clusters are more difficult to detect because of photometric fading. More importantly, they become increasingly rare because of a high mortality rate, the subject of Section~5. The apparent low-mass limit of the clusters is largely an artifact of the cluster definition since objects less massive than, e.g., Orion, become indistinguishable from field stars in galaxies at distances beyond a few Mpc. At the high-mass end, clusters are exceedingly rare. They follow a luminosity function with a power-law index $\sim -2$ (Whitmore 2003), and few are expected on the basis of number statistics. The current record holder is W3 in NGC 7252 whose mass, mass-to-light ratio, and compactness places this super star cluster in an otherwise void region of the fundamental plane (Maraston et al. 2004). In terms of fundamental plane properties, W3 links star clusters and dwarf ellipticals, like M32. 

When comparing synthetic spectra to cluster data, the interpretation may be affected by ``shot noise'' below some cluster mass (Bruzual 2002, Cervi\~no \& Luridiana 2004). Owing to their high luminosity, relatively few massive stars can account for the majority of the cluster light, and small-number statistics comes into play. Cervi\~no \& Luridiana found from Monte-Carlo simulations that shot noise will introduce errors of more than 10\% for cluster masses below $10^4 - 10^5$~\msun, depending on age and wavelength range. For a normal stellar initial mass function (IMF), this corresponds to 200~--~2000 O stars.

Different spectral synthesis models agree quite well among each other when main-sequence stars dominate the light. In Fig.~2 I am comparing the predicted color evolution for a single stellar population from different sources. The older the population and the longer the wavelength, the larger are the discrepancies between the models. These differences are caused by different ingredients and assumptions for the properties of red supergiants (RSGs) and asymptotic giant branch (AGB) stars. All models account only for stellar evolution; no cluster-related effects are included (cf. Section~5). OB stars contribute to the light during the first tens of Myr and only at the shortest wavelengths. The colors are quite insensitive to population variations among massive blue stars. Models at higher resolution predicting line profiles are needed if OB stars are to be tested. Such models exist and have proven to be rather successful (de~Mello et al. 2000, Leitherer et al. 2001).

\section{Impact of Different Physical Ingredients}

\begin{figure}[h]
\begin{center}
\plottwo{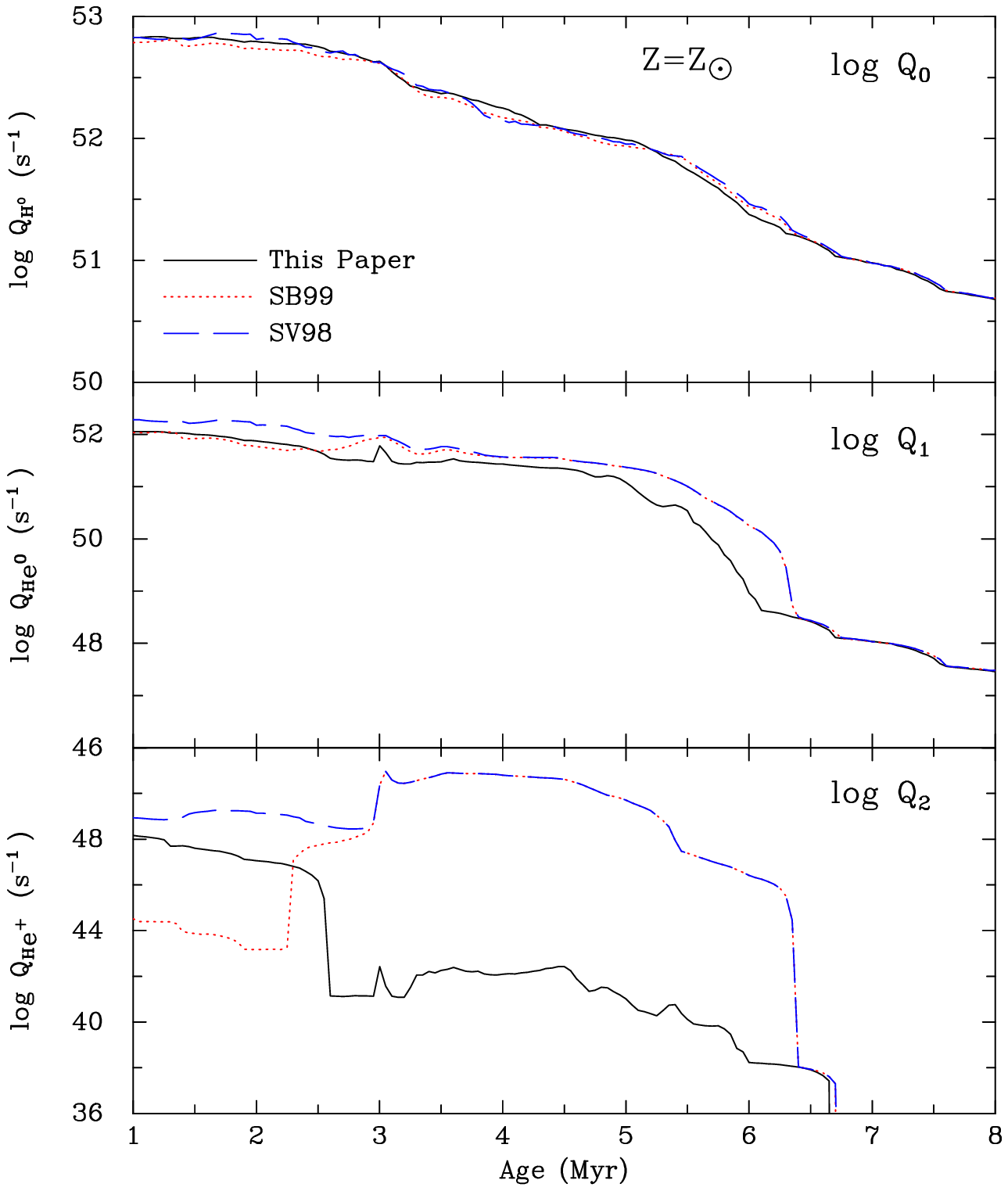}{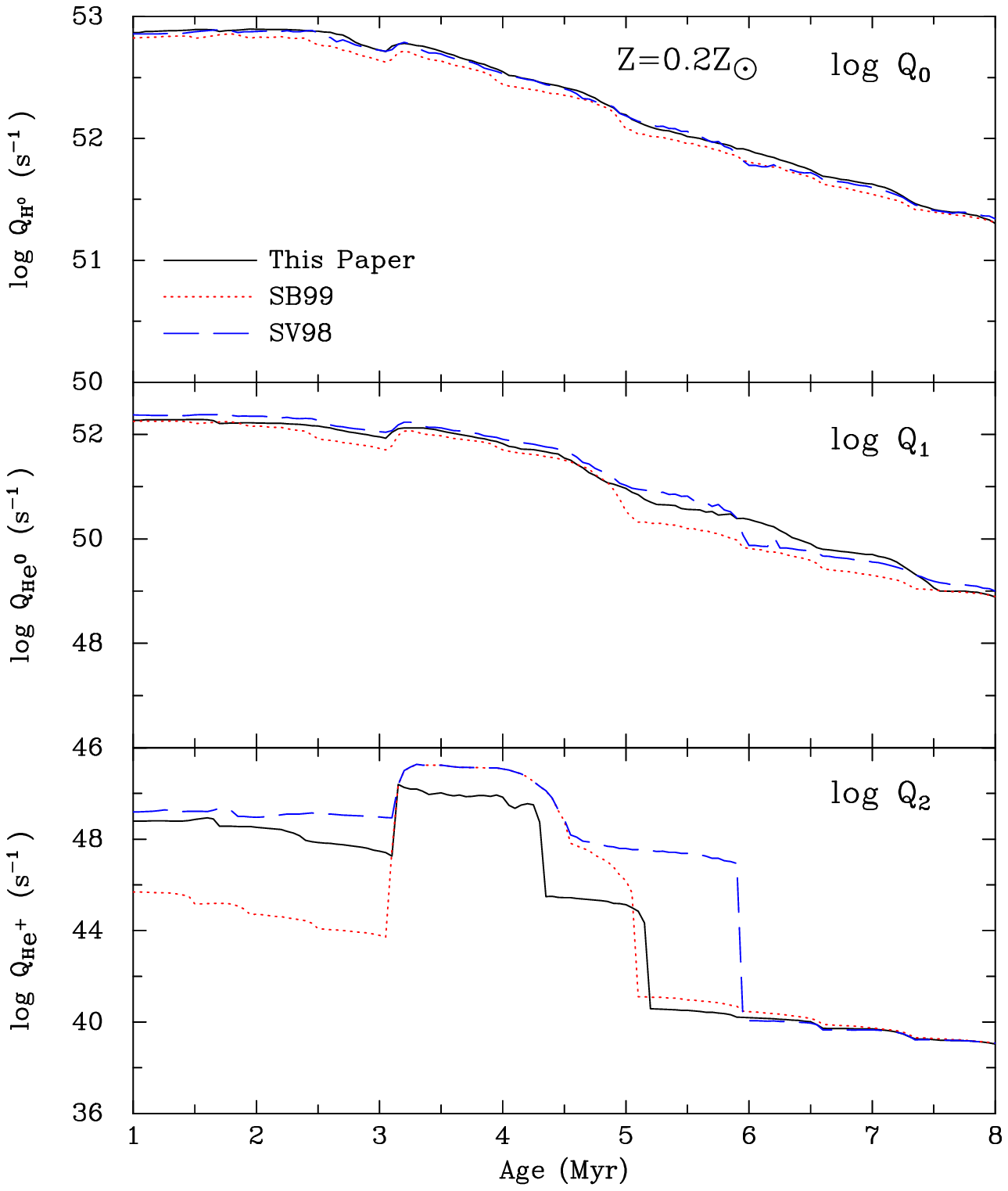}
\end{center}
\caption{The evolution of the photon luminosity for ages of 1~-–~8~Myr in the ionizing continua of hydrogen (top), neutral helium (middle) and ionized helium (bottom) at 0.2 and \zsun\ for an instantaneous burst at time steps of 0.1~Myr. The models of Smith, Norris, \& Crowther 2002 (solid) are compared to the ionizing fluxes of: Leitherer et al. (1999; SB99, dotted) using the stellar library of Lejeune, Cuisinier, \& Buser (1997) and the Schmutz et al. (1992) W-R models; the Schaerer \& Vacca (1998) models (SV98, dashed) using CoStar and Schmutz et al. (1992) W-R atmospheres. From Smith et al. (2002).}
\end{figure}

Recent progress in the field of hot-star atmospheres has led to major revisions of synthesis codes that rely on these atmospheres. Most notably, dynamical non-LTE atmospheres with complete line-blanketing have finally become available. These atmospheres are particularly relevant for the ionizing continua of the hottest stars, such as Wolf-Rayet (W-R) stars. A previous generation of unblanketed W-R atmospheres by Schmutz, Leitherer, \& Gruenwald (1992) was extensively used in the Starburst99 code (Leitherer et al. 1999) but produced a too hard radiation field for metal-rich stars. Smith, Norris, \& Crowther (2002) released a fully blanketed version of such models, combined with a set of O-star atmospheres from WM-basic (Pauldrach, Hoffmann, \& Lennon 2001). These new models are compared to the previous generation in Fig.~3. The differences are negligible in the hydrogen ionizing continuum, moderate for neutral helium, and dramatic in the ionized helium continuum. The changes are caused by two effects: first, blanketing affects the escaping radiation, and second, revised mass-loss rates affect the wind density structure and ionization balance. The coupling between atmospheres and evolution models is not fully self-consistent but must be done empirically, i.e., by calibrating the resultant model predictions with data. Therefore the predictions at the shortest wavelengths should be taken with care.

Dopita et al. (2006) used the Starburst99 code with the Smith et al. (2002) atmosphere as input for photo-ionization modeling with Mappings. Standard line ratios were calculated and compared to an extensive sample of ionizing star clusters. The agreement was quite satisfactory.

The new generation of blanketed O-star atmospheres led to a significant revision of the relation between spectral type and effective temperature \teff. Indications for up to 10\% lower temperatures were found independently in the UV (Bianchi \& Garcia 2002, Crowther et al. 2002) and mid-infrared (Mart\'{\i}n-Hern\'andez et al. 2002). The previous overestimate can be understood as due to the neglect of line-blanketing in the previous models. Blanketing causes additional heating, thereby shifting the helium ionization balance towards higher values. As the spectral types of O stars are defined by the ratio of ionized over neutral helium, a given spectral type will be associated with a lower \teff\ in the new models. Martins, Schaerer, \& Hillier (2005) have published tables for the latest calibration. While the revision is fundamental in many aspects, it is of minor importance for the hydrogen ionizing fluxes predicted by synthesis models, which do not make use of a relation between spectral type and \teff. Rather, these models rely on the relation between stellar ionizing flux and \teff\ directly, which remains essentially unchanged.

The \teff\ scale for cool massive stars in the Galaxy and the Magellanic Clouds has recently been revised by Levesque et al. (2005, 2006). Using the latest version of MARCS atmospheres (Gustafsson et al. 2003) and accounting for the effects of circumstellar dust lowers \teff\ by about 10\%. While this revision is not fully sufficient for bringing the observed and predicted photometric properties of metal-poor RSG populations into agreement (see the following section), it essentially reconciles observed and predicted positions in the Hertzsprung-Russell diagram (HRD) of the solar neighborhood.  

To summarize, I note the significant progress in atmospheric modeling that has translated into vastly improved spectral synthesis modeling. In contrast, stellar evolution modeling for massive stars still poses many challenges and introduces major uncertainties in synthetic models as I will discuss in the next section.

\section{Loose Ends and Future Developments}

One of the pillars of population synthesis are stellar evolution models. Previous modeling of massive star evolution emphasized the importance of the extension of the convective core and of the stellar mass loss, which were thought (together with standard nuclear processing) to be the most relevant ingredients in massive star evolution. This view has changed quite dramatically in recent years. It has become clear that stellar rotation plays a dominant –-- and in some cases even {\em the} dominant –-- role in the evolution of stars with masses above 10~\msun\ (Maeder \& Meynet 2000). The rotation velocity of massive stars may be as high as several hundred km~s$^{-1}$ on the zero-age main-sequence but decreases rapidly when stellar winds are strong because of angular momentum loss. The coupling between rotation and mass loss is complex and not yet fully understood. As a result, the previous generation of stellar evolution models, which does not take into account rotation, may be subject to revision. This in turn will affect all population synthesis modeling relying on these tracks.

A new set of tracks has been released by the Geneva group (Meynet \& Maeder 2005), which addresses stellar rotation. This set of tracks is still incomplete and not meant to fully replace the currently available full set of tracks without rotation. These new tracks, however, are invaluable in exploring the fundamental effects caused by stellar rotation and gauging the impact on stellar population modeling. 
 
The new evolution models with rotation were implemented into Starburst99 by V\'azquez et al. (2006). The resulting color evolution of a stellar cluster is compared to earlier models without rotation in Fig.~4. 
The contribution from RSGs and AGB stars at the longest wavelengths is most obvious in ($V - R$) and ($V - K$). The effects on the colors are delayed with respect to the old models. After 10~--~15~Myr the colors are redder by up to 1.5 magnitudes in the new models with rotation. This effect is due to the redward evolution of stars with masses lower than 30~\msun\ to the RSG phase. In the new models, the stars spend a longer fraction of their life times in the red part of the HRD, rather than evolving back to higher temperatures on blue loops. The peak seen around ${\log}{\rm(Age [Gyr])} \sim -1.4$ in Fig.~5 is the effect of a blue loop at 9~\msun\ that is still present. 

\begin{figure}[h]
\begin{center}
\includegraphics[scale=0.7]{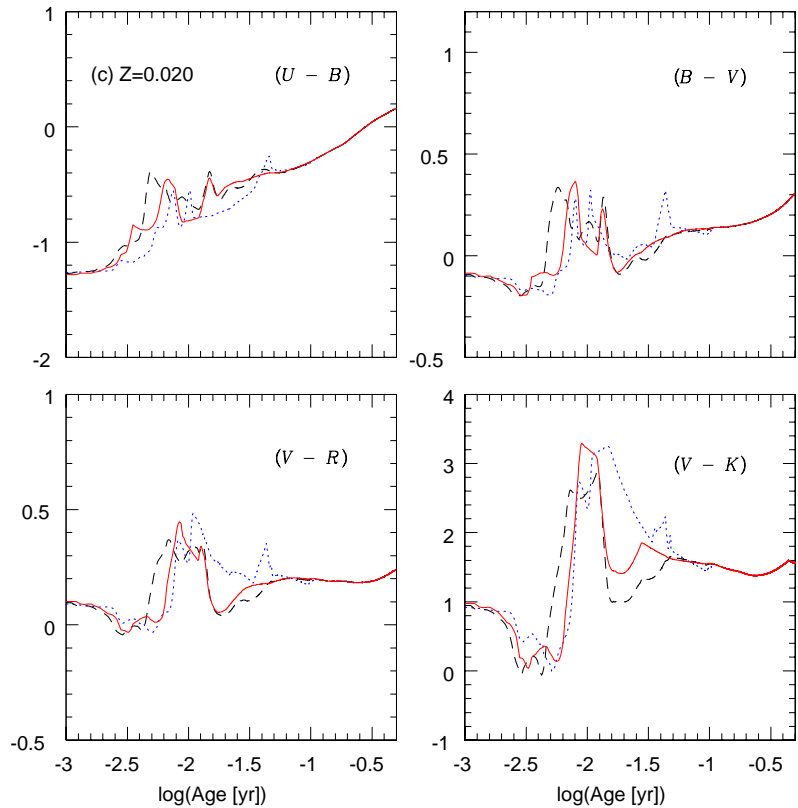}
\end{center}
\caption{Color evolution of a standard single stellar population at \zsun. Solid: Geneva tracks currently used in Starburst99; dashed: new Geneva tracks with revised mass-loss rates and $v_{\rm rot} = 0$~km~s$^{-1}$; dotted: new tracks with revised mass-loss rates and $v_{\rm rot} = 300$~km~s$^{-1}$. From V\'azquez et al. (2006).}
\end{figure}
 
The trend of redder colors due to a stronger RSG contribution is present at all chemical abundances studied. The new models may help alleviate or even solve a long-standing puzzle in population synthesis: the observed colors of metal-poor star clusters in RSG dominated  phased are redder than predicted by previous models (Origlia et al. 1999).

The new models with rotation are more luminous on the main-sequence and in the blue post-main-sequence evolution, including the W-R phase. The most significant consequence is an increased output of ionizing radiation at ages when O and W-R stars are present. Fig.~5 illustrates this point. This figure shows the number of ionizing photons below 912~\AA, 504~\AA, and 228~\AA\ for a young stellar cluster. The photon output of the old and new models is similar during epochs when W-R stars are not present. In contrast, significant differences are observed in the presence of W-R stars between 3 and 8 Myr. The models with rotation generate more ionizing photons than the original tracks. The effect is most pronounced for the ionized He
continuum, which is opaque with the old tracks but becomes more luminous by many orders of magnitude in the new models. The effect is also noticeable in the neutral He continuum, as well as in the neutral hydrogen continuum. The latter displays a factor of 2 increase over the old models. This has immediate consequences for, e.g., the calibration between the ionizing luminosity and the star-formation rate and its dependence on $Z$, as the W-R/O ratio is strongly $Z$ dependent. As a word of caution, V\'azquez et al. (2006) assumed the same coupling between the evolution and atmosphere models for the new models as they did for the old models. This assumption is still untested and needs observational verification.

\begin{figure}[h]
\begin{center}
\includegraphics[scale=0.5]{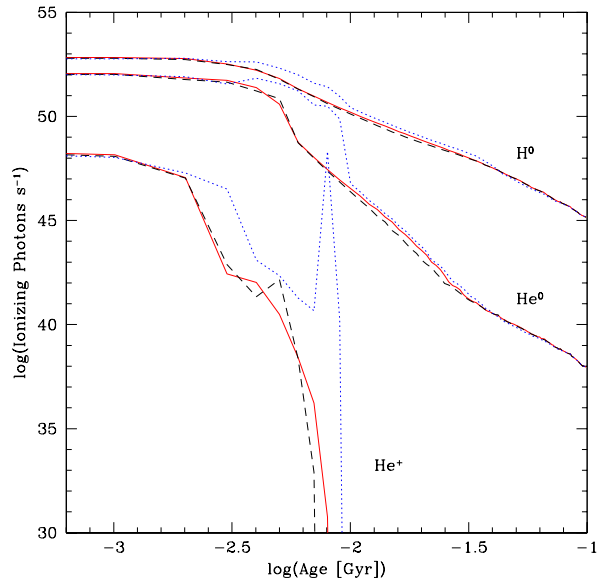}
\end{center}
\caption{Evolution of the ionizing photon output of a standard single stellar population at \zsun. Solid: Geneva tracks currently used in Starburst99; dashed: new Geneva tracks with revised mass-loss rates and $v_{\rm rot} = 0$~km~s$^{-1}$; dotted: new tracks with revised mass-loss rates and $v_{\rm rot} = 300$~km~s$^{-1}$. The three sets of graphs show the results for the stellar continua below 912~\AA, 504~\AA, and 228~\AA. From V\'azquez et al. (2006).}
\end{figure}

\section{Combining Stellar and Cluster Evolution}

Traditionally, synthesis models assume star cluster properties to exclusively result from the superposition of the stellar properties. While this assumption can often lead to useful predictions, it is clear that additional phenomena are relevant and should be taken into account.

\subsection{Mass segregation at birth}

 The formation mechanism of massive stars is still under debate. In the competitive accretion scenario (Bonnnell et al. 2001), stars in a common gravitational potential accrete material from the surrounding gas. Since the gas density is higher in the central region of a star forming cluster, accretion rates are higher in the center. As a result, the most massive stars are more likely located in the cluster center, and the stellar IMF is predicted to be biased towards massive stars in the innermost cluster region. Observationally, this effect would manifest itself in aperture-size effects: a smaller aperture would favor a flatter IMF, i.e., one biased towards massive stars. Synthesis models must make an assumption for the IMF, and spatial variations would introduce an additional free parameter. Stolte et al. (2005) reported evidence for a very top-heavy IMF in the central 0.4 pc of the Arches Cluster near the Galactic Center. Their deep adoptive optics near-IR photometry suggest a turn-over of the IMF and a deficit of stars below 6~\msun.

\subsection{Infant mortality}

NGC~4038/39 (``The Antennae'') is the closest example of a merging galaxy pair. The system has been undergoing a strong starburst for the past $\sim$10$^8$~yr, which led to the formation of a plethora of young star clusters (Whitmore et al. 2005). The typical cluster ages are of order $10^7$~yr, which is at least an order of magnitude less than the starburst time scale. In this case, the age distribution of the clusters immediately reflects their survival rate. Fall, Chandar, \& Whitmore (2005) determined the luminosities and ages of about $10^4$ clusters in the Antennae from broad- and narrowband photometry and compared them to synthesis models (Fig.~6). Even after correcting for completeness and contamination by stars Fig.~6 suggest a significant deficit of clusters older than about 10 to 20~Myr in comparison with the models. There are as many clusters with ages less than 10~Myr as there are with ages between 10~Myr and 1~Gyr.

\begin{figure}[h]
\begin{center}
\includegraphics[scale=0.5]{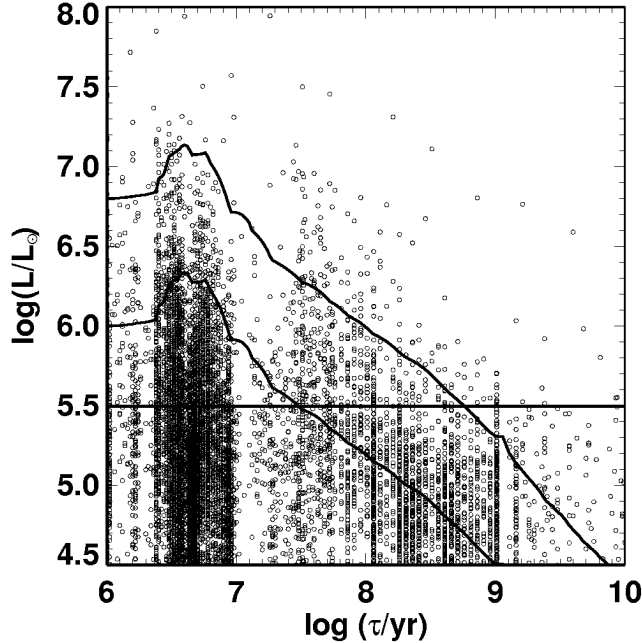}
\end{center}
\caption{Luminosity versus age of star clusters in the Antennae galaxies. The sloping graphs indicate the predicted photometric decline of clusters with initial masses of $3\times10^4$~\msun\ and $2\times10^5$~\msun. 
The horizontal line at $L = 3\times10^5$~\lsun\ shows the upper limit for stellar contamination.
The vertical gap at 10 to 20~Myr is an artifact of the age-fitting procedure. From Fall, Chandar, \& Whitmore (2005).}
\end{figure}

The deficit of older clusters is interpreted as due to ``infant mortality'', i.e., the destruction of clusters shortly after birth. The combined effects of ionizing radiation, stellar winds, and supernovae disrupts and expels the intracluster medium. As a result, the stars quickly become gravitationally unbound, similar to the case of a Galactic OB association.

\subsection{Cluster mass loss due to stellar evolution}

Massive stars continuously lose mass during their evolution, first via stellar winds and shell ejections and later via supernova explosions. A single stellar population with a standard IMF is predicted to have lost close to 30\% of the original stellar mass by the age of 20 Myr (Leitherer et al. 1999). The gas is initial injected into the cluster interstellar medium but will become gravitationally unbound and most likely be expelled from the star cluster. This has important consequences for synthesis modeling. An often employed technique for determining the low-mass end of the IMF are velocity dispersion measurements (Larsen, Brodie, \& Hunter 2005). If the system is virialized (among other conditions that need to be met), the velocity dispersion is a measure of the total mass, which can then be compared to predicted mass-to-light ratios as a function of age. Traditional synthesis models account only for the photometric fading. Inclusion of the mass decrease of the cluster significantly alters the predicted mass-to-light ratio (Smith \& Gallagher 2001).

\subsection{Mass segregation and loss of low-mass stars}

Simulations of the dynamical evolution of star clusters predict that massive stars will migrate towards the cluster center, whereas low-mass stars are preferentially located in the outer regions. This ``mass segregation'' is commonly observed in populous clusters in the Milky Way and in the Magellanic Clouds. As a result of mass segregation, low-mass stars are more likely to become unbound and escape from the gravitational field of the star cluster. Obviously, the removal of a particular stellar species will affect the photometric properties of the star cluster, yet previous synthesis models of the photometric evolution have all ignored this effect.

\begin{figure}[h]
\begin{center}
\includegraphics[scale=0.7]{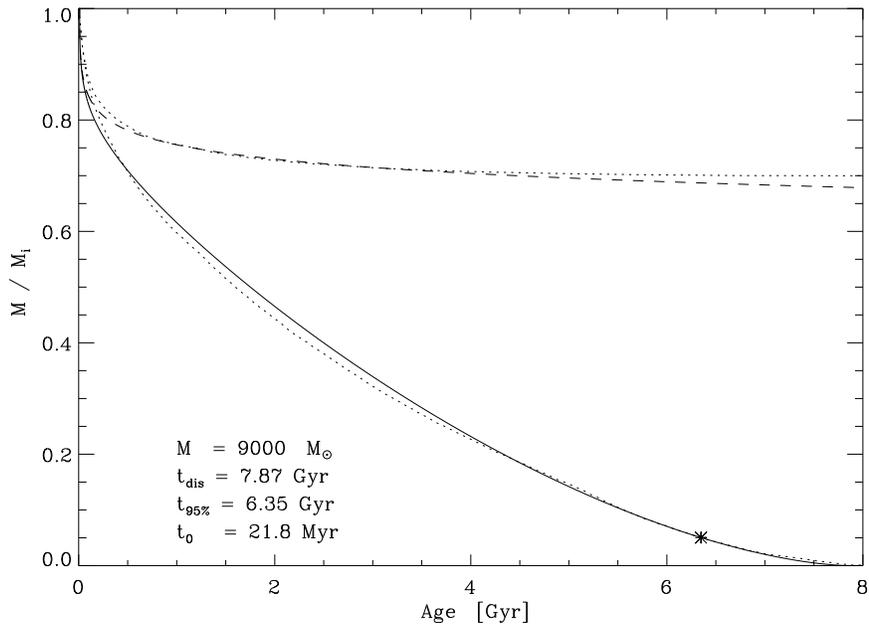}
\end{center}
\caption{Evolution of the mass of a cluster with an initial mass of
  9000 \msun. The solid line is the decreasing mass due to stellar evolution
  and loss of low-mass stars; the dashed line shows the decreasing mass if stellar
  evolution were the only mass-loss mechanism. The dotted lines indicate the 
  results of more elaborate N-body simulations. The asterisk indicates the moment when 95\% of the
  initial mass is lost. From Lamers, Anders, \& de Grijs (2006).}
\end{figure}

Recently, Lamers, Anders, \& de Grijs (2006) modeled the photometric evolution of a star cluster undergoing mass loss due to stellar evolution (discussed in Section~5.3) and the loss of low-mass stars due to mass segregation. Their results are illustrated in Fig.~7. Mass segregation induced cluster mass loss is negligible during the first $\sim$100~Myr when stellar winds and supernovae completely dominate. Later-on, mass segregation and loss of low-mass stars sets in, and the cluster mass decreases by 95\% at an age of $\sim$6.3~Gyr. Lamers et al. found that the color evolution is the same as that predicted for standard models without loss of low-mass stars during the first 40\% of the cluster evolution.  Between 40 and 80\% of the lifetime the cluster becomes slightly bluer than predicted by models without stellar ejections because the cluster has lost a large fraction of red low-mass stars. The most dramatic changes occur between about 80 and 100\% of the total age. Then AGB stars and red giants are the dominant contributors to the red colors. These stars are much redder than stars at the low-mass end of the main-sequence. Removing the low-mass  stars via cluster mass loss will enhance the color contrast between the remaining main-sequence and the red evolved population, thereby making the cluster colors redder. This reddening will result in an {\em overestimate} of the cluster age from broad-band photometry if standard cluster evolution models are used. The effect is substantial and can lead to an overestimate by a factor of $\sim$4 near the end of the cluster's life. Charlot et al. (1993) pointed out a similar modification of the cluster colors caused by an IMF truncated at the low-mass end rather than by mass segregation and stellar ejection.

\subsection{Tidal disruption}

Eventually all star clusters, except for the most massive ones, may be dissolved by tidal disruption in the gravitation field of the host galaxy. The disruption time depends on internal cluster properties, such as total mass and IMF, density, and velocity dispersion, as well as on external conditions, such as the location in the galaxy and the amount of tidal heating. The time scale may be as long as a Hubble time for the most massive clusters in the galaxy periphery or comparable to the evolutionary time scale of massive stars for a cluster close to the galactic center (Lamers, Gieles, \& Portegies Zwart 2005).

\acknowledgements 
I would like to thank the organizers for the opportunity to be part of this special meeting. My special thanks go to Henny Lamers for the privilege of being his colleague, collaborator, and friend.


\begin{thebibliography}{}

\bibitem[]{h5} Bianchi, L., \& Garcia, M.\ 2002, \apj, 581, 610 

\bibitem[]{x9} Bonnell, I.~A., Clarke, C.~J., Bate, M.~R., \& Pringle, J.~E.\ 2001, \mnras, 324, 573 

\bibitem[]{m8} Bruzual, G. A. 2002, in Extragalactic Star Clusters, IAU Symp. 207, eds. D. Geisler, E. K. Grebel, \& D. Minniti (San Francisco: ASP), 616

\bibitem[]{d5} Bruzual, G., \& Charlot, S. 2003, \mnras, 344, 1000

\bibitem[]{a6} Cervi{\~n}o, M., \& Luridiana, V.\ 2004, \aap, 413, 145 

\bibitem[]{s2}  Cervi\~no, M., Mas-Hesse, J. M., \& Kunth, D. 2002, \aap, 392, 19 

\bibitem[]{j4} Charlot, S., Ferrari, F., Mathews, G.~J., \& Silk, J.\ 1993, \apjl, 419, L57 

\bibitem[]{d8} Crowther, P.~A., Hillier, D.~J., Evans, C.~J., Fullerton, A.~W., De Marco, O., \& Willis, 
A.~J.\ 2002, \apj, 579, 774 

\bibitem[]{b2} de Mello, D.~F., Leitherer, C., \& Heckman, T.~M.\ 2000, \apj, 530, 251 

\bibitem[]{j7} Dopita, M. A., et al. 2006, \apj, in press (astro-ph/0608062)

\bibitem[]{r3} Fall, S.~M., Chandar, R., \& Whitmore, B.~C.\ 2005, \apjl, 631, L133 

\bibitem[]{a2} Fioc, M., \& Rocca-Volmerange, B. 1997, \aap, 326, 950 

\bibitem[]{c2} Gonz\'alez Delgado, R.~M., Cervi{\~n}o, M., Martins, L.~P., Leitherer, C., \& Hauschildt, P.~H.\ 2005, 
\mnras, 357, 945 

\bibitem[]{v7} Grebel, E.~K., \& Chu, Y.-H.\ 2000, \aj, 119, 787 

\bibitem[]{a0} Gustafsson, B., Edvardsson, B., Eriksson, K., Mizuno-Wiedner, M., J{\o}rgensen, U.~G., \& 
Plez, B.\ 2003, in Stellar Atmosphere Modeling, eds. I. Hubeny, D. Mihalas, \& Klaus Werner (San Francisco: ASP), 331 

\bibitem[]{k8} Holtzman, J.~A., et al.\ 1992, \aj, 103, 691 

\bibitem[]{x5} Lada, C.~J., \& Lada, E.~A.\ 2003, \araa, 41, 57 

\bibitem[]{m1} Lamers, H.~J.~G.~L.~M., Anders, P., \& de Grijs, R.\ 2006, \aap, 452, 131 

\bibitem[]{c9} Lamers, H.~J.~G.~L.~M., Gieles, M., \& Portegies Zwart, S.~F.\ 2005, \aap, 429, 173 

\bibitem[]{z9} Larsen, S.~S., Brodie, J.~P., \& Hunter, D.~A.\ 2004, \aj, 128, 2295 

\bibitem[]{f3} Leitherer, C., Le{\~a}o, J.~R.~S., Heckman, T.~M., Lennon, D.~J., Pettini, M., \& Robert, 
C.\ 2001, \apj, 550, 724 

\bibitem[]{a3} Leitherer, C., Schaerer, D., Goldader, J. D., Gonz\'alez Delgado, R. M., Robert, C., Foo Kune,
    D., de Mello, D., Devost, D., \& Heckman, T. M. 1999, \apjs, 123, 3

\bibitem[]{g4}  Lejeune, T., Cuisinier, F., \& Buser, R.\ 1997, \aaps, 125, 229 

\bibitem[]{q6} Levesque, E.~M., Massey, P., Olsen, K.~A.~G., Plez, B., Josselin, E., Maeder, A., \& Meynet, 
G.\ 2005, \apj, 628, 973 

\bibitem[]{b1} Levesque, E.~M., Massey, P., Olsen, K.~A.~G., Plez, B., Meynet, G., \& Maeder, A.\ 2006, 
\apj, 645, 1102 

\bibitem[]{m5} Maeder, A., \& Meynet, G.\ 2000, \araa, 38, 143 

\bibitem[]{n6} Maraston, C.\ 1998, \mnras, 300, 872 

\bibitem[]{f9} Maraston, C., Bastian, N., Saglia, R.~P., Kissler-Patig, M., Schweizer, F., \& Goudfrooij, P.\ 
2004, \aap, 416, 467 

\bibitem[]{z1} Mart{\'{\i}}n-Hern{\'a}ndez, N.~L., Vermeij, R., Tielens, A.~G.~G.~M., van 
der Hulst, J.~M., \& Peeters, E.\ 2002, \aap, 389, 286 

\bibitem[]{l2}  Martins, F., Schaerer, D., \& Hillier, D.~J.\ 2005, \aap, 436, 1049 

\bibitem[]{s7} Meynet, G., \& Maeder, A.\ 2005, \aap, 429, 581

\bibitem[]{x3} Origlia, L., Goldader, J.~D., Leitherer, C., Schaerer, D., \& Oliva, E.\ 1999, \apj, 514, 96  

\bibitem[]{s4} Pauldrach, A.~W.~A., Hoffmann, T.~L., \& Lennon, M.\ 2001, \aap, 375, 161 

\bibitem[]{j3} Robert, C., Pellerin, A., Aloisi, A., Leitherer, C., Hoopes, C., \& Heckman, T. M. 2003, \apjs, 144, 21

\bibitem[]{z2} Schaerer, D., \& Vacca, W.~D.\ 1998, \apj, 497, 618 

\bibitem[]{b9} Schmutz, W., Leitherer, C., \& Gruenwald, R.\ 1992, \pasp, 104, 1164 

\bibitem[]{r5} Schulz, J., Fritze-v. Alvensleben, U., M\"oller, C. S., \& Fricke, K. J. 2002, \aap, 392, 1

\bibitem[]{h8} Smith, L.~J., \& Gallagher, J.~S.\ 2001, \mnras, 326, 1027 
 
\bibitem[]{c6} Smith, L.~J., Norris, R.~P.~F., \& Crowther, P.~A.\ 2002, \mnras, 337, 1309 

\bibitem[]{f8} Steidel, C.~C., Adelberger, K.~L., Shapley, A.~E., Pettini, M., Dickinson, M., \& 
Giavalisco, M.\ 2003, \apj, 592, 728 

\bibitem[]{l6} Stolte, A., Brandner, W., Grebel, E.~K., Lenzen, R., \& Lagrange, A.-M.\ 2005, \apjl, 628, L113 

\bibitem[]{k6} V{\'a}zquez, G.~A., Carigi, L., \& Gonz{\'a}lez, J.~J.\ 2003, \aap, 400, 31 

\bibitem[]{i9} V\'azquez, G. A., \& Leitherer, C. 2005, \apj, 621, 695

\bibitem[]{k7} V\'azquez, G. A., Leitherer, C., Schaerer, D., Meynet, G., \& Maeder, A. 2006, \apj (submitted)

\bibitem[]{a1} Walborn, N. R., \& Barb\'a, R. H. 1999, in New Views of the Magellanic Clouds, IAU Symp. 190, eds. Y.-H. Chu, N. Suntzeff, J. Hesser, \& D. Bohlender (Dordrecht: Kluwer), 252

\bibitem[]{d6} Whitmore, B. C. 2003, in A Decade of Hubble Space Telescope Science, eds. M. Livio, K. Noll, \& M. Stiavelli (Cambridge:CUP), 153

\bibitem[]{f4} Whitmore, B.~C., et al.\ 2005, \aj, 130, 2104 

\end{thebibliography}
\end{document}